# EVASÃO ESCOLAR: UMA DURA REALIDADE
# (SCHOOL EVASION: A HARD REALITY)

**Valessa Leal Lessa de Sá Pinto**[1]
**Frederico Alan de Oliveira Cruz**[2]
Universidade do Grande Rio
Mestrado Profissional em Ensino das Ciências na Educação Básica

***Resumo:*** O presente trabalho teve como objetivo mostrar o perfil dos estudantes que abandonaram os estudos em uma escola de Ensino Médio, situada na cidade de São João de Meriti, localizado no estado de Rio de Janeiro, por meio da análise estatística. Os índices apresentados retratam uma realidade indesejável com a evasão de quase 20% dos alunos inicialmente matriculados, além de mostrar que mais da metade dos alunos encontra-se em séries distorcidas em relação a idade.
***Palavras-chave:*** Evasão Escolar, Ensino Médio, Estatística Educacional.

***Abstract:*** The present work has as objective to show the profile of students who abandoned the studies in a High School, located in São João de Meriti city, municipal district of Rio de Janeiro state, by means of statistical analysis. The presented indices portray an undesirable reality with almost 20% school evasion, beyond showing that more the half of the students not standing in adequate series.

## 1. Introdução

Um dos graves problemas da educação pública brasileira refere-se a evasão escolar. O tema que exige muita reflexão e discussão e, aguarda alternativas de solução em âmbito da nacional, está, segundo dados apresentados pelo Instituto Brasileiro de Geografia e Estatística (IBGE), atingindo taxas preocupantes em toda a educação básica Brasileira (Estado de São Paulo, 2008). Nesse estudo, foi mostrado que 5% das crianças entre 7 e 14 anos abandonam a escola e que se considerarmos os adolescentes, com idades de 15 à 17 anos, a evasão sobe para quase 20% dos alunos inicialmente matriculados na escolas, isto é, são quase dois milhões de alunos fora da escola.

A evasão desses jovens está ligada a muitos obstáculos, muitas vezes intransponíveis, para permanecerem na escola e concluírem seus estudos. A necessidade de trabalhar, convívio com a violência, problemas familiares, dificuldade de acesso à escola e a má qualidade de ensino, entre outros motivos, são os fatores mais comuns para o abandono escolar por esses jovens, isto é, a evasão está relacionada não apenas a escola mas também à família, às políticas de governo em uma determinada região e ao próprio aluno. Esse conjunto de situações faz com que o aluno deixe de acreditar que a escola proporcionará a ele um futuro melhor, visto que a educação que eles recebem no presente é precária em relação ao conteúdo, a formação de valores, ao preparo para o mundo do trabalho. Em outras palavras, como não há garantias de que está aprendendo algo realmente modificador o aluno acaba saindo da escola, como foi constatado em recente pesquisa da Fundação Getúlio Vargas, onde muitos estudantes afirmaram não querer mais estudar pois a escola que está aí não os atrai (Jornal Gazeta do Povo, 2007).

Outro motivo para esse desinteresse está relacionado as sucessivas reprovações sofridas pelo aluno e que acabam por gerar mais um elemento desmotivador. Dessa forma é muito comum vincular abandono escolar, na maioria das vezes, a uma histórico de repetências. A retenção, sem dúvida, tem um peso significativo na decisão de continuar ou não os estudos, pois na maioria das vezes, a repetência, é seguida pelo abandono escolar (Lopes e Menezes, 2002; Garschagen e Belchior, 2007), mas esse não é o único fator responsável por esse fenômeno e deve-se tomar cuidado com propostas de não reprovação nas séries fundamentais, como o elemento salvador da educação. É verdade que em alguns países a reprovação é proibida pura e simplesmente (Garschagen e Belchior, 2007), caso de Noruega e Suécia entre tantos outros, mas devemos refletir também sobre a construção cultural desses países e como seus cidadãos encaram a educação.

---

[1] valessaleal@bol.com.br
[2] frederico.cruz@unigranrio.edu.br

Voltando a realidade brasileira, os alunos que apesar das repetências, nas séries do ensino fundamental, chegam ao Ensino Médio, o fazem com idade avançada, gerando uma distorção em relação a série que ele frequenta e a sua idade. Em algumas situações, a distorção série-idade chega a 42% entre jovens de 18 a 24 anos, isto é, esses alunos estão em séries adequadas para outras faixas etárias, devido à evasão e/ou repetências (Rigotto e Souza, 2005).

Devido a distorção série-idade, o aluno, em muitos casos, acaba escolhendo as turmas de Ensino Médio noturno na tentativa de conclusão do ensino básico, pois apesar de toda dificuldade encontrada nas etapas do ensino esse aluno ainda busca na escola igualdade de oportunidade e através dela formas de não-exclusão (Togni e Soares, 2007). Essa suposta solução esbarra, na maioria das vezes, em um ensino que adota as mesmas metodologias e em alguns casos a mesma carga horária mas sem o mesmo rigor que é adotado nas turmas regulares do período diurno (Santos, 2006), isto é, esse alunos ficam renegados a um ensino na maioria das vezes de baixa qualidade e de pouca ou nenhuma utilidade.

## 2. Perfil Escolar e Metodologia

Considerando todas as questões acima, nosso trabalho teve objetivo fundamental determinar a taxa de evasão de uma escola de Ensino Médio do Rio de Janeiro. A escola escolhida localiza-se na região central do município de São João de Meriti, região da Baixada Fluminense, e pertence a rede pública estadual. Essa escolha ocorreu por ela estar localizada em um município que faz fronteira com outros municípios da Baixada Fluminense e ser de fácil acesso dos moradores de alguns bairros da periferia do município do Rio de Janeiro.

Em 2007, havia no CIEP 169, três diretores, duas coordenadoras pedagógicas, duas orientadoras educacionais, cinco funcionários de secretaria, onze funcionários de apoio, limpeza e merenda, três funcionários de biblioteca, uma agente pessoal, seis inspetores e setenta e seis professores.

A estrutura física da escola apresenta boas condições de uso, com os espaços sempre limpos e arrumados. Os alunos contam com uma quadra de esportes, laboratório de informática com computadores que podem ser usados para trabalhos escolares, sala de vídeo, auditório, biblioteca com um acervo razoável, e os devidos materiais de cada local, bem conservados.

Após a observação da estrutura escolar, partimos para conhecer a clientela do CIEP 169 – Maria Augusta Correia no ano considerado. A partir da consulta do Livro de Matrícula da escola constatamos que foram matriculados 2408 alunos, sendo que 1424 desses alunos foram matriculados em turmas do Ensino Médio e os outros 984 em turmas do segundo segmento do Ensino Fundamental, que compreende as séries do sexto ao nono ano.

Os alunos do Ensino Médio da escola foram distribuídos em trinta e três turmas, somando-se os turnos da manhã e noite, enquanto que os alunos do Ensino Fundamental estavam distribuídos em vinte uma turmas, nos turnos tarde e noite. Após esse mapeamento inicial decidimos que o nosso público alvo seriam os alunos matriculados na turmas do Ensino Médio, visto todos os questionamentos apresentados na introdução desse artigo.

O passo seguinte a determinação do grupo de alunos do nosso trabalho, foi realizada uma consulta na Ata de Resultados Finais da escola, a fim de coletar quantos alunos, inicialmente matriculados na escola, tinham evadido. Após determinarmos o total de alunos evadidos e não evadidos, realizamos um estudo relacionando os alunos com os seus respectivos municípios de origem e assim determinar quais eram as áreas atendidas pela escola.

Com a identificação dos alunos evadidos, determinamos o valor percentual em relação ao total dos alunos evadidos da escola, em função do município de origem e relativamente ao próprio município.

Após a determinação do percentual de alunos evadidos por município, nosso estudo analisou a evasão em relação as séries e os turnos em que a escola funciona. Essa análise foi realizada para confirmamos dois questionamentos realizados pelos professores, onde o primeiro deles é se há uma grande evasão de alunos do turno da noite e se essa evasão ocorre principalmente na primeira série.

A análise posterior foi realizada nos dados dos homens e mulheres escritos na escola, com o objetivo de verificar se existia uma maior tendência de abandono de um desses grupos ou se existia um equilíbrio entre eles.

A determinação sobre o perfil de idade também foi realizado. Inicialmente verificamos as idades de todos os alunos que cursaram o Ensino Médio no ano de 2007. O estudo aqui foi importante para sabermos se o aluno atendido pela escola já passou por repetências ou desistências. Os dados obtidos foram então separados em classes, que foram determinadas não por métodos estatísticos convencionais e sim baseados no perfil do aluno que frequenta o Ensino Médio. A faixa etária adequada para esse segmento de ensino é de 15 a 17 anos. Sendo assim, consideramos que seria conveniente a separação em classes de quatro anos e não através de métodos estatísticos.

Para determinarmos a idade com maior frequência de evasão, fizemos uma análise estatística entre as medidas de tendência central, verificando qual a mais apropriada para representar a idade crítica de evasão.

## 3. Resultados e Discussão

O total de alunos evadidos, 288, e não evadidos, que chamamos de presentes, foi de 1136. Esse resultado mostrou que o número de evadidos representou 20,2% de todos os alunos matriculados nas trinta e três turmas da escola (Figura 1).

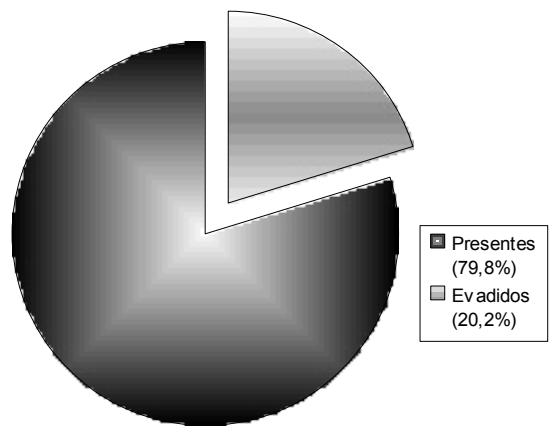

Figura 1: Percentual de alunos presentes e evadidos da escola

Esse primeiro dado já se mostrou preocupante não apenas pelo número em si, mas também por que acreditávamos que esse número fosse menor que o obtido.

Em relação a origem dos alunos da escola, foi possível mostrar que além de atender alunos do próprio município, a escola recebe alunos de quase todos os municípios que fazem limite com São João de Meriti. As exceções são Nilópolis e Mesquita. Segundo a diretora pedagógica da escola, o número de atendimentos a esses lugares é proporcional à precariedade das suas escolas nessas regiões devido a falta de políticas públicas voltadas à educação. No caso do município de Nilópolis existe uma parceria com as escolas da rede estadual presentes nesse município. O município de São João de Meriti, SJM, apresentou um total de 691 alunos presentes e 194 evadidos, para o município de Belford Roxo foram contabilizados 322 alunos presentes e 72 evadidos, no município do Rio de Janeiro encontramos 111 alunos presentes e 21 evadidos, em Duque de Caxias são 10 presentes e apenas 1 evadido e finalmente Nova Iguaçu com 2 alunos presentes e nenhum evadido (Fig 2).

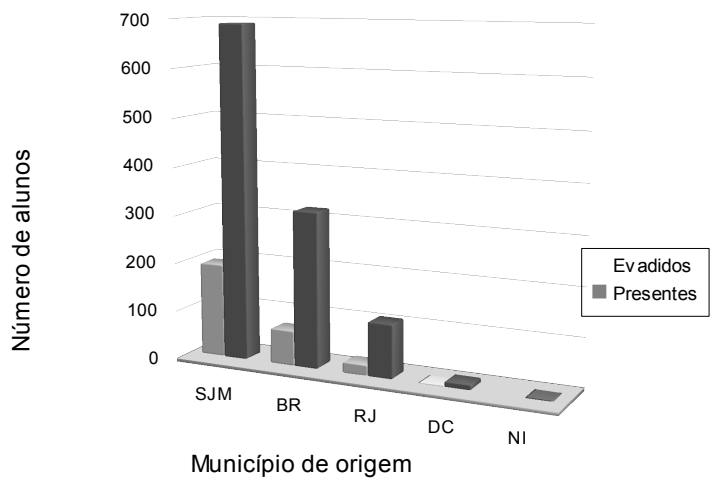

Figura 2: Alunos evadidos e presentes em relação ao município de origem

O problema da evasão é maior para os alunos com origem em SJM, onde o CIEP se localiza. No entanto isso já era esperado, pois é a cidade da maioria dos alunos (figura 3).

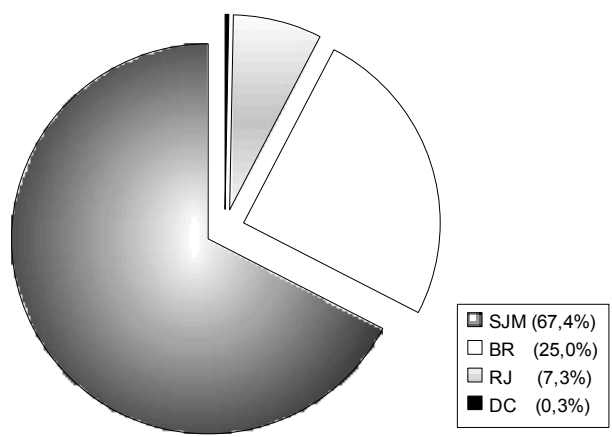

Figura 3: Percentual de alunos evadidos em função da origem

Quando comparamos o total de alunos evadidos em função de matrículas, para moradores do mesmo município, vemos que ocorre um equilíbrio relativo entre três regiões: São João de Meriti, Rio de Janeiro e Belford Roxo. Para São João de Meriti temos 21,9% de evasão do total de alunos inicialmente matriculados, 18,3% para Belford Roxo, 15,9% para o Rio de Janeiro. Para os alunos com origem nos municípios de Duque de Caxias e Nova Iguaçu, a evasão relativa foi de 9,1% e 0%, respectivamente.

A diferença dos valores percentuais, desses dois últimos municípios em relação aos outros, ocorre devido ao número menor de alunos oriundos desses municípios e assim afetar de alguma maneira esses dados.

A distribuição dos alunos evadidos do Ensino Médio, segundo a série e o turno (Fig. 4), nos mostrou que existe um grande número de alunos evadidos na primeira série e que essa evasão diminui a cada série, tanto para os alunos do turno da manhã como para os alunos do curso noturno. A explicação para tal fenômeno estaria ligada a muitos fatores como jornada trabalho-estudo. A evasão menor na terceira série deve-se ao fato do aluno estar no último ano e estar próximo do completar seu objetivo que é a conclusão do ensino médio.

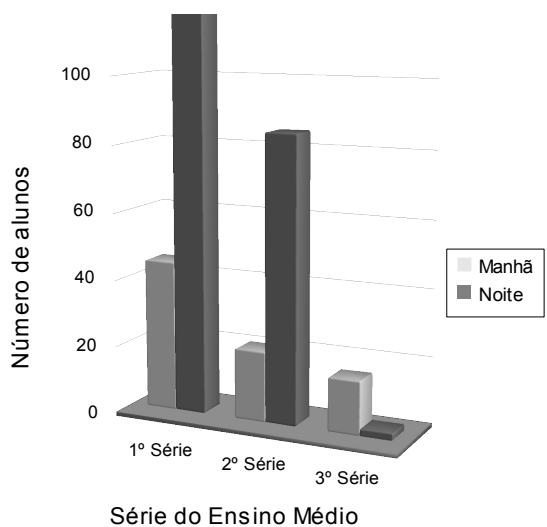

Figura 4: Número de alunos evadidos nas séries em cada turno

Os dados coletados nos mostraram que a evasão noturna corresponde a praticamente 72% da evasão total ocorrida (Fig. 5a), sendo que as turmas de primeira série, como um todo, contribuem com quase 57% da evasão da escola (Fig. 5b).

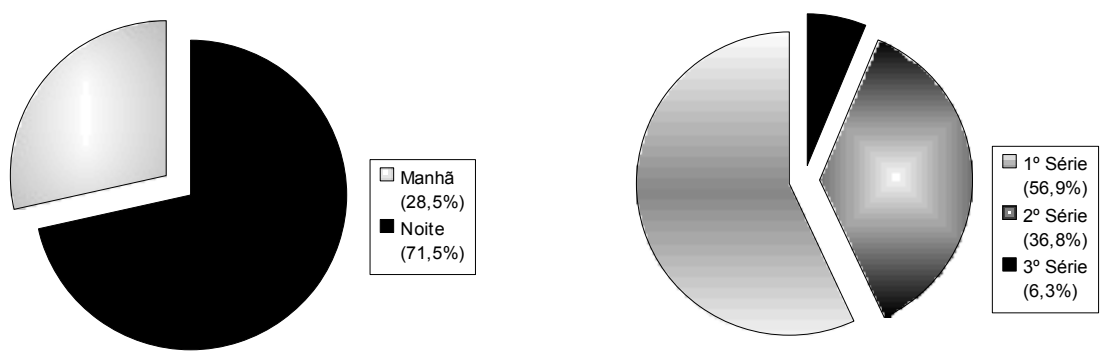

Figura 5a: Distribuição percentual de evadidos por turno.   Figura 5b: Distribuição percentual de evadidos por série

O abandono no turno da noite é justificado por grande parte dos alunos, pelo cansaço devido ao trabalho e não pela incompatibilidade de horário com o serviço, como no turno da manhã. O desânimo em relação aos estudos é marcante no período da noite. Dando ênfase às séries, temos uma quantidade alta de evadidos na 1ªs e 2ªs séries. A evasão escolar nestas etapas é caracterizada pelo tempo de estudo que ainda falta para a conclusão do curso, baseado em algum dos argumentos já comentados. Assim, encontramos na 3ª série os "sobreviventes" dessa longa caminhada, ainda com algumas desistências. Essa falta de interesse, provavelmente, também é determinada por outros fatores que citamos no início do texto, mas que dificilmente são dados como justificativa para as orientadoras da escola.

Em relação a distribuição de evadidos por sexo houve um equilíbrio, com uma evasão maior para alunos do sexo masculino. Essa diferença não consideramos significativa, pois a maioria dos alunos da escola são homens (Fig 6), apesar de alguns trabalhos já mencionarem que existe um maior abandono de alunos do sexo masculino (Carvalho, 2004).

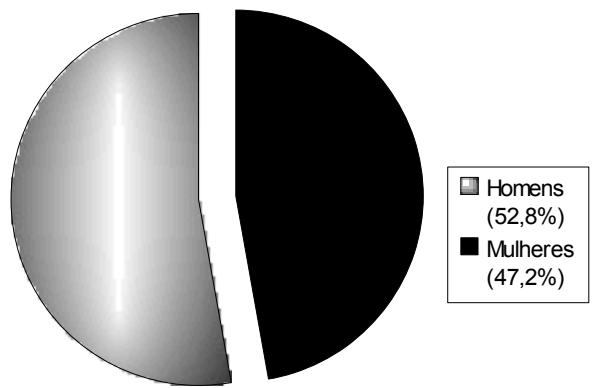

Figura 6: Distribuição dos alunos evadidos por sexo

Além dos motivos comuns entre ambos os sexos para a saída da escola, como necessidade de trabalho, outras razões podem influir na decisão, tais como: alistamento militar para os rapazes e gravidez para as moças. Esse último contribui muito para evasão pois aproximadamente 20% das alunas nessa situação abandonam os estudos (Fanelli et al. 2002).

A distribuição das idades dos alunos presentes na escola está representada na Fig. 7 e mostra dados interessantes, a primeira delas é que aproximadamente 10% dos alunos da escola estavam acima dos 26 anos e quase 60% estavam em séries distorcidas em relação a idade.

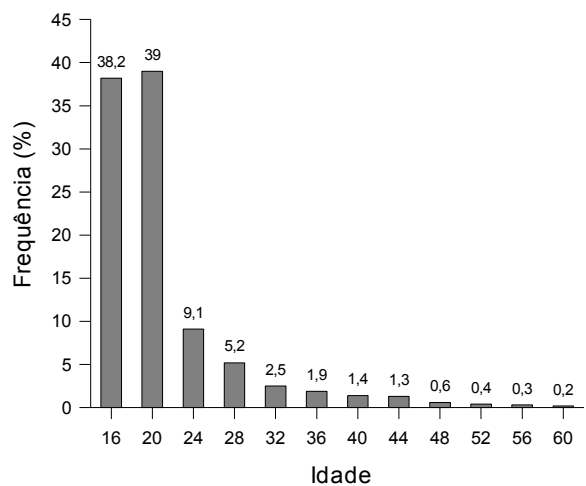

Figura 7: Distribuição da idade por classes

Na distribuição da idade dos alunos evadidos, a faixa etária de 18 a 22 anos possui aproximadamente 45% de evasão escolar, muito em função do grande número de alunos com essa idade presentes na escola (Fig 8).

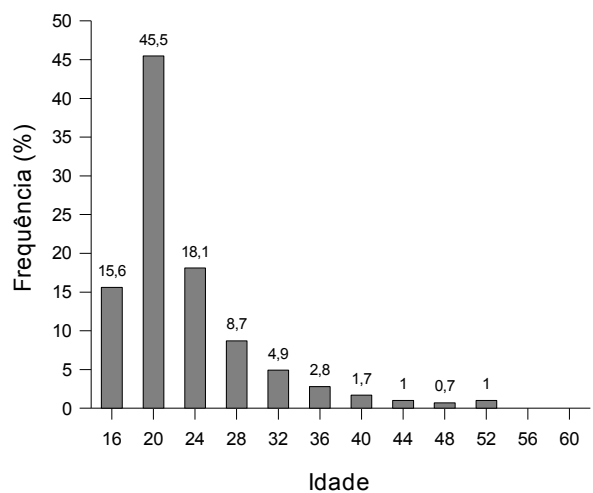

Figura 8: Distribuição da evasão em função da idade

Como não dispúnhamos de uma distribuição simétrica, nossa observação pousou sobre a mediana, pois pela sua definição essa medida de tendência central divide a distribuição em dois grupos iguais (Duquia e Bastos, 2006) e parecia mais conveniente dentro do nosso estudo do que usarmos a média aritmética que podia ser influenciada pelos valores extremos. O valor da mediana ficou em torno de 23 anos, indicando que alunos entre 22 e 26 anos são as idades críticas de evasão na escola. Esse valor é coincidente com o percentual relativo do número de alunos matriculados e evadidos. No caso dos alunos com idade entre 22 e 26 anos 40% deles evadiram da escola, maior percentual relativo entre as classes. Em relação a classe modal é fácil perceber que ela está entre os alunos entre 18 e 22 anos, mas essa classe representa o maior número de alunos presentes nessa escola e isso pode influenciar nessa medida.

## 4. Conclusões e Considerações Finais

Os resultados obtidos na pesquisa, infelizmente, confirmaram nossas suposições em relação à evasão escolar. Na escola estudada houve uma taxa de 20% de evasão, em relação aos alunos inicialmente matriculados, e que coincide com os índices apresentados pelo IBGE. Outro ponto interessante obtido desses dados foi mostrar que aproximadamente 60% dos alunos encontram-se fora da idade escolar ideal, indicando uma tendência de repetências e/ou abandonos já ocorridos durante a vida escolar desses alunos, isto é, constatamos que a escola tem cerca de 40% do total de alunos do Ensino Médio com a faixa etária adequada, ou seja, dentro do intervalo de 14 a 18 anos.

Foi possível observar de forma direta que o atual modelo de "*escola para todos*" precisa passar por mudanças no intuito de reduzir o alto índice de desistência, principalmente no turno da noite, e ser um espaço de ampliação de conhecimentos, interação social e qualificação profissional. A escola deverá planejar novas estratégias de combate à evasão, olhando não para os alunos em idade adequada, mas sim para os alunos que estão fora dessa faixa etária. Dessa forma, esperamos que esta análise estatística sirva, de fato, para o diagnóstico de dificuldades enfrentadas na escola e que incentive a prática de ações a favor de uma educação de qualidade.

## 5. Agradecimentos

A Direção, professores e funcionários do CIEP 169 pela colaboração e apoio durante o desenvolvimento desse trabalho e ao Prof. Francisco Antônio Lopes Laudares pelas suas sugestões durante a fase de escrita desse artigo.

## 6. Referências Bibliográficas

__________Ata de resultados finais do Ensino Médio. CIEP 169: Maria Augusta Correia, Secretaria Estadual de Educação, São João de Meriti: 2007.

CARVALHO, M.P. (2004): "O fracasso escolar de meninos e meninas: articulações entre gênero e cor/raça*", in *Cadernos Pagu*, n.22, p.247-290.

DUQUIA, R.P; BASTOS, J.D. (2006): "Medidas de tendência central: Onde a maior parte dos indivíduos se encontra?", in *Scientia Medica*, n.16, v.4, p. 190-194.

FANELLI, C.M.T., FERMAN, E., PEREIRA, J.L. PEREIRA, R.C.R., RIOS, S.P.S. (2002): "A gravidez e a evasão escolar: uma questão de consciência cidadã", in *Anais do I Congresso Brasileiro de Extensão Universitária*, v.1, p.1-7.

GARSCHAGEN, S., BELCHIOR, F. (2007): "O dilema da repetência e da evasão", in *Desafios*, n.4, v.36, p.34-43.

__________Jornal Gazeta do Povo (2007): "Não querer estudar é o principal motivo da evasão escolar", <http://canais.ondarpc.com.br/gazetadopovo/educacao/conteudo.phtml?id=650265> [Consulta: set. 2008]

__________Jornal O Estado de São Paulo (2008): "Evasão escolar cresce entre beneficiados do Bolsa-Família", <http://www.estadao.com.br/nacional/not_nac136993,0.htm> [Consulta: set. 2008].

__________Livro de Matrícula, CIEP 169: Maria Augusta Correia, Secretaria Estadual de Educação, São João de Meriti: 2007.

LOPEZ, F. L.; MENEZES, N.A. (2002): "Reprovação, avanço e evasão escolar no Brasil", in *Pesquisa e Planejamento Econômico*, n.32, v.3, p.417-451.

RIGOTTO, M.E., SOUZA, N.J. (2005): "Evolução da Educação no Brasil, 1970-2003", in *Análise*, v.16, n.2, p. 339-358.

SANTOS, M.E.M. (2006): "Ensino Médio noturno: do proposto ao realizado", in *Anais do IV Encontro de Pesquisa em Educação da UFPI*, v.1, p.1-13.

TOGNI, A.C., SOARES, M.J. (2007): "A escola noturna de Ensino Médio no Brasil", in *Revista Iberoamericana de Educación*, n.44, p.61-76.